\documentclass[preprint,12pt]{elsarticle}

\usepackage{graphicx}

\usepackage{amssymb}

\journal{Physics of the Dark Universe}

\begin{document}

\begin{frontmatter}



\title{The importance of local measurements for cosmology}


\author[1,2]{Licia Verde}
\author[1,2]{Raul Jimenez}
\author[3]{Stephen Feeney}
\address[1]{ICREA, and ICC-UB-IEEC, University of Barcelona, Marti i franques 1, 08034, Barcelona, Spain.}
\address[2]{Theory Group, Physics Department, CERN, CH-1211, Geneva 23, Switzerland.}
\address[3]{Department of Physics and Astronomy, University College London, London WC1E 6BT, U.K.}
\begin{abstract}

We explore how local, cosmology-independent measurements of the Hubble constant and the age of the Universe help to provide a powerful consistency check of the currently favored cosmological model (flat $\Lambda$CDM) and model-independent constraints on cosmology. We use cosmic microwave background (CMB) data to define the model-dependent cosmological parameters, and add local measurements to assess consistency and determine whether extensions to the model are justified. At current precision, there is no significant tension between the locally measured Hubble constant and age of the Universe (with errors of 3\% and 5\% respectively) and the corresponding parameters derived from the CMB. However, if errors on the local measurements could be decreased by a factor of two, one could decisively conclude if there is tension or not. We also compare the local and CMB data assuming simple extensions of the flat, $\Lambda$CDM model (including curvature, dark energy with a constant equation of state parameter not equal to  $-1$, non-zero neutrino masses and a non-standard number of neutrino families) and find no need for these extra parameters; in particular, we constrain the effective number of neutrino species to be $N_{\rm eff}<4$ at 95\% confidence. We show that local measurements provide constraints on the curvature and equation of state of dark energy nearly orthogonal to those of the CMB. We argue that cosmology-independent measurements of local quantities at the percent level would be very useful to explore cosmology in a model-independent way.
\end{abstract}


\end{frontmatter}


\section{Introduction}
\label{sec:intro}
Most cosmological constraints and statements are model-dependent.  For example, CMB observations  predominantly probe the  physics of the early Universe, up to a redshift of  $z\sim 1100$ or $\sim 380,000$ years after the big bang.  These observations are then interpreted in terms of %
cosmological parameters  which are defined at $z=0$ (such as physical matter density $\Omega_m h^2$, curvature $\Omega_k$, dark energy density $\Omega_{\Lambda}$, age of the universe, $t_U$ or  Hubble constant $H_0$)  by extrapolation within a given model.

There is therefore an immense added value in measuring  if possible, at least some of these parameters locally, in a way that is independent of the cosmological model. Direct measurements of the Hubble constant  have long been a workhorse to reduce CMB-parameters degeneracies.
However, the  comparison  of low and high redshift measurements is not just useful to break parameters degeneracies (which, again, must be done within a cosmological model) but also to test the underlying model itself. 
Given that, in the currently favored model, 96\% of the Universe is dark and the properties of the elusive dark energy are also not known, it is important to test  directly the model itself. This has been extensively done, for example, by extending the baseline model with one (or more) extra parameter and constraining it. To go beyond parameters fitting and into model testing,  in the Bayesian framework, Bayesian Evidence is often used. Examples of deviations from the baseline model which have been widely explored in the literature include (but are not limited to), adding curvature,  allowing the dark energy to deviate from a cosmological constant in a parameterised way, adding extra  (effective) neutrino species (the Standard Model has 3 families), allowing neutrinos to have non-zero mass, allowing a coupling between dark matter and dark energy, etc.

However one could attempt to make less model-dependent statements. One could ask: is there any indication that  the standard (flat $\Lambda$CDM) cosmological model  is inadequate or incomplete? A natural way to do that is to compare high redshift (e.g., CMB) constraints extrapolated to present time within that model, with direct  measurements at $z=0$. If the underlying model was incorrect there would be no reason for the two approaches to agree.

Measuring the local expansion rate of the  Universe has played a crucial role in observational cosmology since the early days when the expansion of the universe was discovered by Slipher and Hubble \cite{Slipher}. 
Estimates of the age of the Universe were also crucial in defining the cosmological model: the age of the Universe can be obtained from the ages of the oldest objects, and accurate dating of globular clusters has been subject of active investigation for decades.

The importance of these two measurements  is apparent if we write the expansion rate for a cosmology where the  dark energy field is non-negligible:\footnote{Radiation has been neglected here as it does not  appreciably affect  the age.}
\begin{equation}
t_U H_0= \int^\infty_0 \frac{1}{\sqrt{\Omega_m (1+z)^3 + \Omega_{\Lambda}  (1+z)^{3(1-w)} +\Omega_k(1+z)^2}}
\label{eq:exp}
\end{equation}
where $w$ is the  equation of state parameter for the dark energy  $p/ \rho = w$ assumed constant.

It is apparent from the above equation that an accurate determination of $H_0$ and $t_U$ provides a means to constraining composition  of the Universe. Further, they provide the means of checking for a particular model of the Universe by looking at consistency of Eq.~(\ref{eq:exp}) when combined with other measurements of the right-hand side of the equation. Historically, before the discovery of dark energy, the above equation was, within the framework of an Einstein-deSitter model:
\begin{equation}  
t_U = (2/3)/H_0
\label{eq:eds}
\end{equation}
thus accurate measurements of both could have provided a confirmation (or not) of the cosmological model.

Surprisingly, since the early 1950s, Eq.~(\ref{eq:eds}) was never verified by data \cite{agejim,agerev}, with values of $H_0$ in the range $50-80$ km s$^{-1}$ Mpc$^{-1}$ and values for the ages of the oldest stars in the range $13-16$ Gyr. This ``age crisis" (the universe being younger than the objects its contains)  should have been a clear indication for abandoning the  Einstein-deSitter model \cite{BolteHogan}.
Instead the blame was (at least before the Supernovae results of 1999) put on systematic errors associated to the fact that measurements of $H_0$ and $t_U$ required the use of stellar astrophysics and distance determinations.

Today in principle, there is no reason why one cannot obtain accurate estimates of both $H_0$ and $t_U$ by doing accurate observations that reduce significantly  the dependence on stellar modelling (measurement of parallaxes has greatly improved the distance measures for nearby objects). In fact, as we report below, this is already the case for $H_0$ (measured at 3\% accuracy) and $t_U$ (measured at 5\% accuracy).  The accuracy and precision of $H_0$ measurement is  due to the combined effort of the  Hubble Key project and its follow up projects\footnote{http://www.ipac.caltech.edu/H0kp/}\cite{key} and, most recently, the S$H_0$ES project, where observations and analyses were carefully planned to control and reduce systematic errors in building the cosmic distance ladder and keep them below the statistical ones \cite{Riess/etal:2011, Freedman/etal:2012}. 
At present, as far as we know, there is no similar  systematic program  for $t_U$: the local estimate of  the age of the Universe is  obtained from the ages of the oldest stars  and current error are systematic dominated.  

In what follows we explore the implications of local measurements of $H_0$ and $t_U$  for cosmology. We  also speculate what would be the implications  of a reduction of the current errors: while $H_0$  could be measured to 2\% (which is the goal of the SH$_0$ES project \cite{Riess/etal:2011}), we discuss how and how much  the age errors could  be conceivably reduced and the implications for cosmology.

\section{Data}
\label{sec:data}
For the local measurements we use the most recent measurements of the Hubble constant from \cite{Riess/etal:2011} and \cite{Freedman/etal:2012} and the constant on the  age of the Universe obtained from the age determination of  the star HD 140283 \cite{Bond}.
\subsection{Hubble parameter data}

Ref. \cite{Riess/etal:2011} reported an estimate of the Hubble  constant with a 3.3\% error: $H_0=73.8 \pm 2.4$ Km s$^{-1}$Mpc$^{-1}$. Which include both statistical and systematic errors. This estimate includes three different approaches to calibrate the distance ladder (one bases on masers, one on Milky Way  cepheids and one on Large Magellanic Cloud cepheids), but the reported uncertainty is the larger afforded by any two of these.  In this measurement, statistical errors dominate the systematic ones.
To be overzealous, their data actually give a Hubble parameter  measurement  at an effective redshift of 0.04 which is then extrapolated to $z=0$ using a standard $\Lambda$CDM cosmology. This assumption induces a systematic error for non-$\Lambda$CDM  models  which is much smaller and totally negligible compared to the reported error-bars. This is therefore ignored here.

Ref. \cite{Freedman/etal:2012}  reported  an estimate of  the Hubble  constant with a  2.8\% error, which is systematic:  $H_0=73.8 \pm 2.4$ Km s$^{-1}$Mpc$^{-1}$. This is based on a Cepheids-calibrated distance scale. 

The two measurements are remarkably close  to each other, offering high confidence in the result. In addition, by being so similar, the choice of one or the other would not change (qualitatively or quantitatively)  anything in  the following analysis.   However,  for not choosing one over the other we decide to report her and use a ``world average".
In our world average, the  central value is  given by variance-weighted mean (using a straight mean would not have changed the result significantly) and error conservatively  given by the average of the errors: $H_0=74.08 \pm 2.25$ Km s$^{-1}$Mpc$^{-1}$. Note that the error of the mean would have been much smaller ($1.6$ km s$^{-1}$Mpc$^{-1}$); the two measurements are somewhat independent but the  Ref.~\cite{Freedman/etal:2012} errors are systematic (making the error of the mean not appropriate).

\subsection{Stellar ages}
The age of the oldest stars can be used as a proxy for the age of the Universe. On the one hand, obviously,  the Universe can't be younger than the objects it contains. On the other hand in most (if not all) cosmologies,  these objects form very shortly after the big-bang. The  uncertainty over formation time is much smaller --at least at present-- than the  measurement errors. Estimating stellar ages for the Universe has been done traditionally from dating globular clusters, the cooling sequence of white dwarfs or the ages of well measured individual stars near the main sequence turn-off (see e.g. \cite{agejim,agerev,chabkrauss}). Until very recently, none of these techniques had achieved an accuracy below 10\% when including systematic errors.

Recently it has become possible to obtain accurate distances to nearby sub-giants using direct parallax measurements with the Hubble Space Telescope. The star with the best age determination is HD 140283 \cite{Bond}. HD 140283 is a sub giant moving just off the main sequence, thus its luminosity is extremely sensitive to its age. By measuring a very precise parallax to the star with an error of 0.14 mas, the authors were able to reduce the error associated with distance in determining ages of stars to only  $0.3$ Gyr, i.e. only 2\% of the star age. Further, because both the effective temperature and the individual abundances of different chemical elements ($O$ and $Fe$) were measured, accurate stellar isochrones could be used to obtain an accurate age: $14.46 \pm 0.3$ (statistical  only) $\pm 0.8$ (statistical +systematic) Gyr, an estimate with a 5\% accuracy (including systematic errors).

HD 140283 has low metallicity but not primordial, $[Fe/H] = -2.4$ and $[O/H] = -1.67$. Note that for these low-metallicity stars, oxygen abundance is the main uncertainty when modelling their evolution. Therefore our adopted error is dominated by systematic uncertainty. In \S \ref{sec:future} we speculate how the error could be significantly reduced.

The formation of such low-metallicty stars is uncertain within current models that constraint the formation of the first generation of stars (PopIII) and second generation (popII) (e.g., \cite{Loeb}), but it is common among the different models to point to a redshift range $z \approx 20-30$. This corresponds to an age after the Big-Bang of $0.1-0.2$ Gyr. This uncertainty is much smaller than current errors. In our analysis we convert the age of the star to the age of the Universe by adding to the star age $0.15 \pm 0.05$ Gyr assuming a Gaussian  distribution (cutting negative tails when appropriate).  We obtain the following estimate for the age of the Universe: $t_U=14.61 \pm 0.8$.

\subsection{High redshift (CMB) data}

Despite the imminent release of the Planck\footnote{http://sci.esa.int/science-e/www/area/index.cfm?fareaid=17} results (and data), we argue that the main  point of the present paper, that is the added value of combining (cosmology-dependent) primary CMB and  (cosmology-independent) local measurements, will not be changed by the new Planck  data. We therefore proceed using the WMAP 9 years data (WMAP9)  \cite{wmap9a, wmap9b}. In some cases we will use WMAP 7 years data (WMAP7) \cite{wmap7a,wmap7b}.

\section{Consistency \& Cosmological implications}
\label{sec:cosmo}
\begin{figure}[h]
\hspace*{-0.5cm}
                 \includegraphics[scale=0.4]{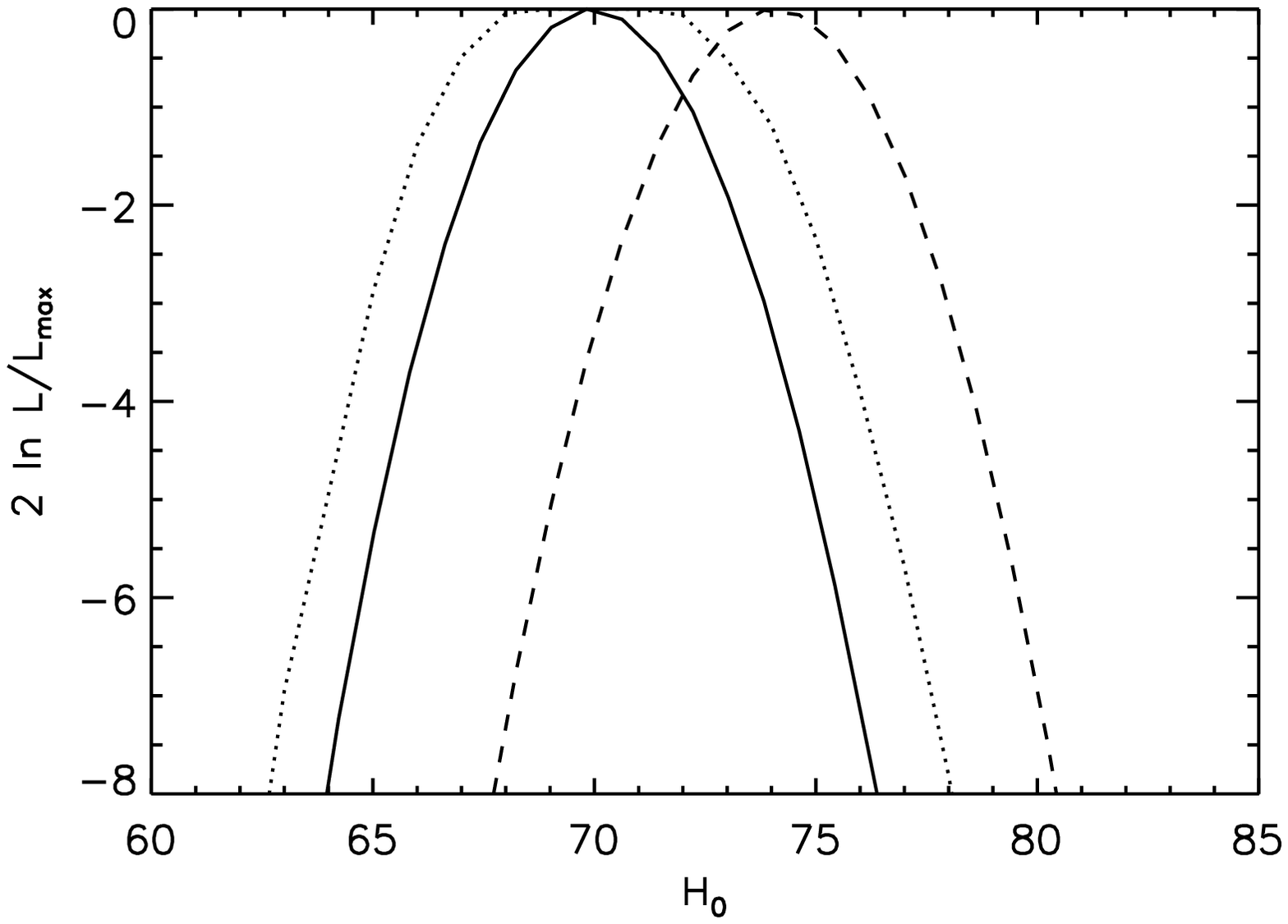}
                 \includegraphics[scale=0.4]{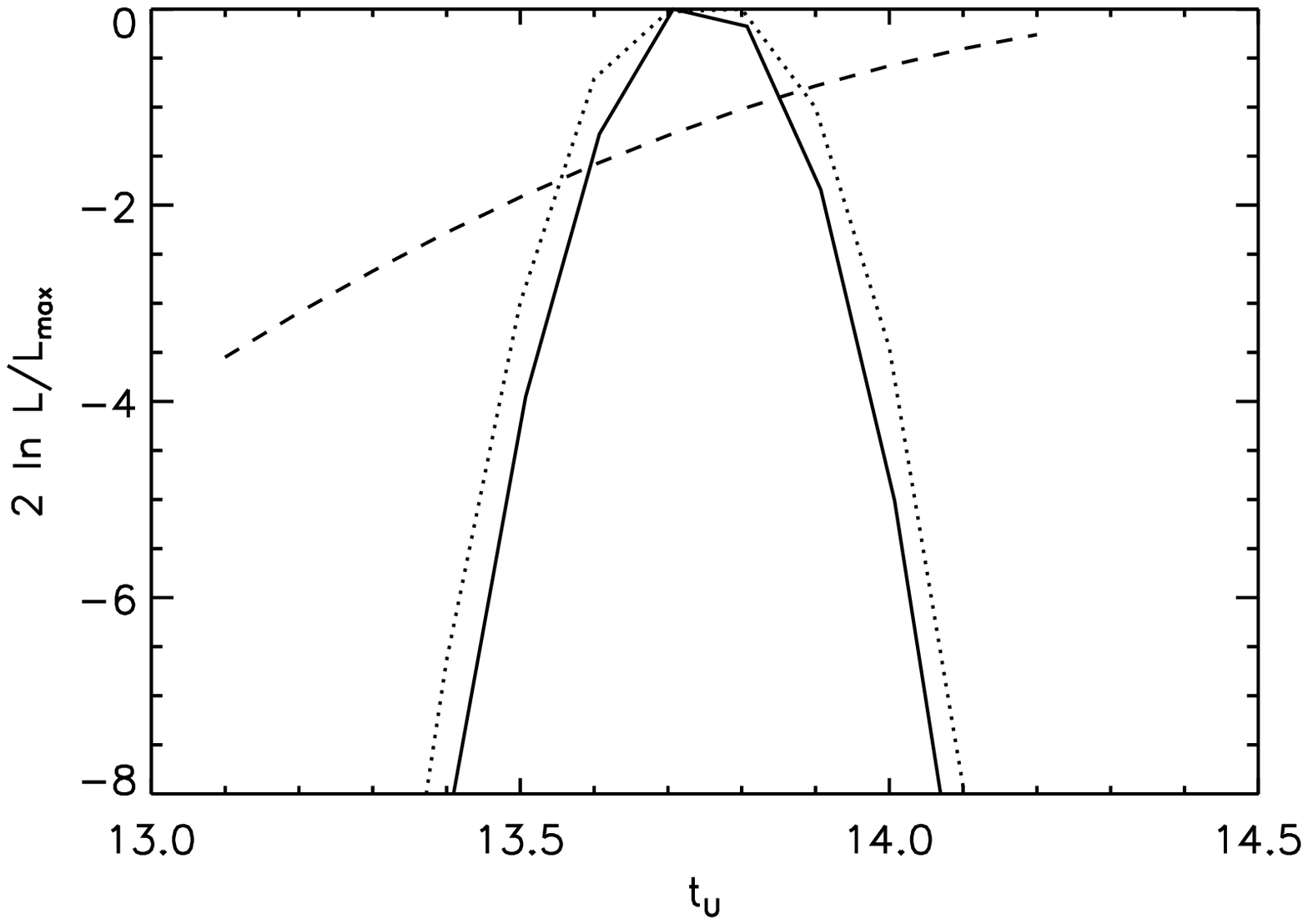}  
        \caption{comparison of WMAP9 constraints and direct $z=0$ measurements: Left panel: $H_0$, right panel: age of the Universe $t_U$. In all panels: solid corresponds to WMAP9 posterior distribution, dotted to WMAP9 profile likelihood and dashed to the chi-square for the $z=0$ measurement. \label{fig:consistency}
}    
\end{figure}

\begin{figure}[h]
\centering
                \includegraphics[scale=0.7]{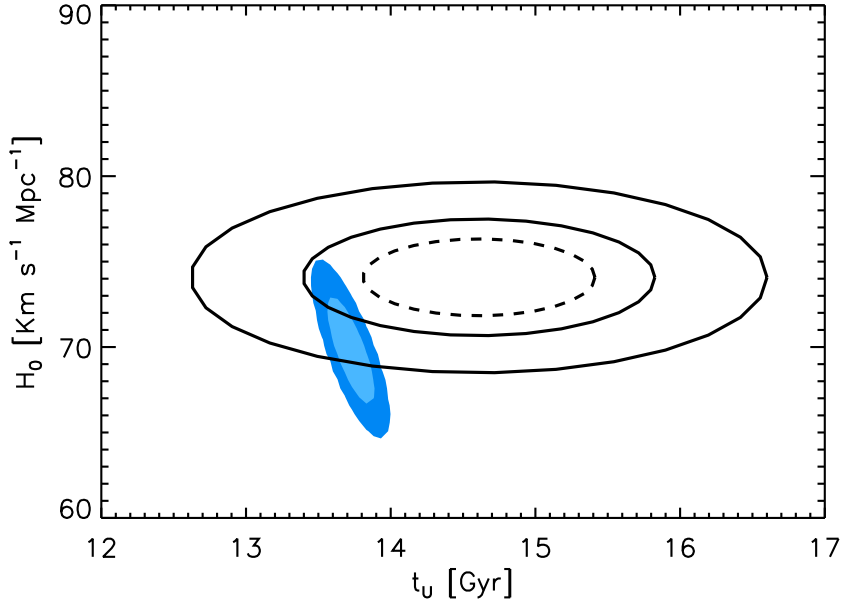}
        \caption{Comparison of WMAP9 constraints  in the $t_U$--$H_0$ plane for a flat $\Lambda$CDM model (filled contours) and direct $z=0$ measurements (transparent). For the local measurements, the dashed line shows the 1-$\sigma$ (one parameter) and the solid lines show the 1 and 2 $\sigma$ joint  confidence regions.}
          \label{fig:2dcomparison}
\end{figure}

We  first explore the CMB-extrapolated constraints on $H_0$ and $t_U$ with the direct constraints above within the flat $\Lambda$CDM baseline model. The CMB analyses are carried out in the Bayesian framework and the reported constraints are therefore prior-dependent. For completeness we also report prior-independent constraints based on the profile likelihood ratio e.g.,\cite{Reid, alma, feeney}. The comparison is shown  in Figs. \ref{fig:consistency} and \ref{fig:2dcomparison}. Fig. \ref{fig:consistency} shows $H_0$ on the  left panel  and age on the right panel. In order to show all curves on the same figure we plot the quantity $2 \ln L/L_{max}$ where $L$ denotes  the marginalized posterior, the profile likelihood or $-\chi^2$ for WMAP9 or  $z=0$ measurements.
\footnote{In passing, it is interesting to note that even in the simple flat $\Lambda$CDM model the WMAP9 constraints on $H_0$ (and age) are  in part affected by the adopted prior (the dotted --based on likelihood values-- and solid --based on posterior-- lines are different). }

The 2-dimensional representation of Fig.~(\ref{fig:2dcomparison}) is   however more illustrative.

In fact one can appreciate how tight the CMB constraints are (within this model) but also that there might be a hint of tension between the high redshift and local measurements which is much less evident when considering 1 dimensional constraints. 

We do not find this hint statistically significant: the {\it joint} 1-$\sigma$ confidence regions overlap (even if by little).  One must however be careful when combining measurements that are not fully consistent as the resulting error-bars could be unnaturally small. On the other hand there may be extensions of the baseline model that bring the two datasets in better agreement.

 Given that the  $t_U$ error is systematic dominated,  better understanding of these errors   (see discussion in \S \ref{sec:future}) could shed light on this hint of tension.

We next consider (parameterized) deviations from this model.

\begin{figure}[h]
\centering
                \includegraphics[scale=0.55]{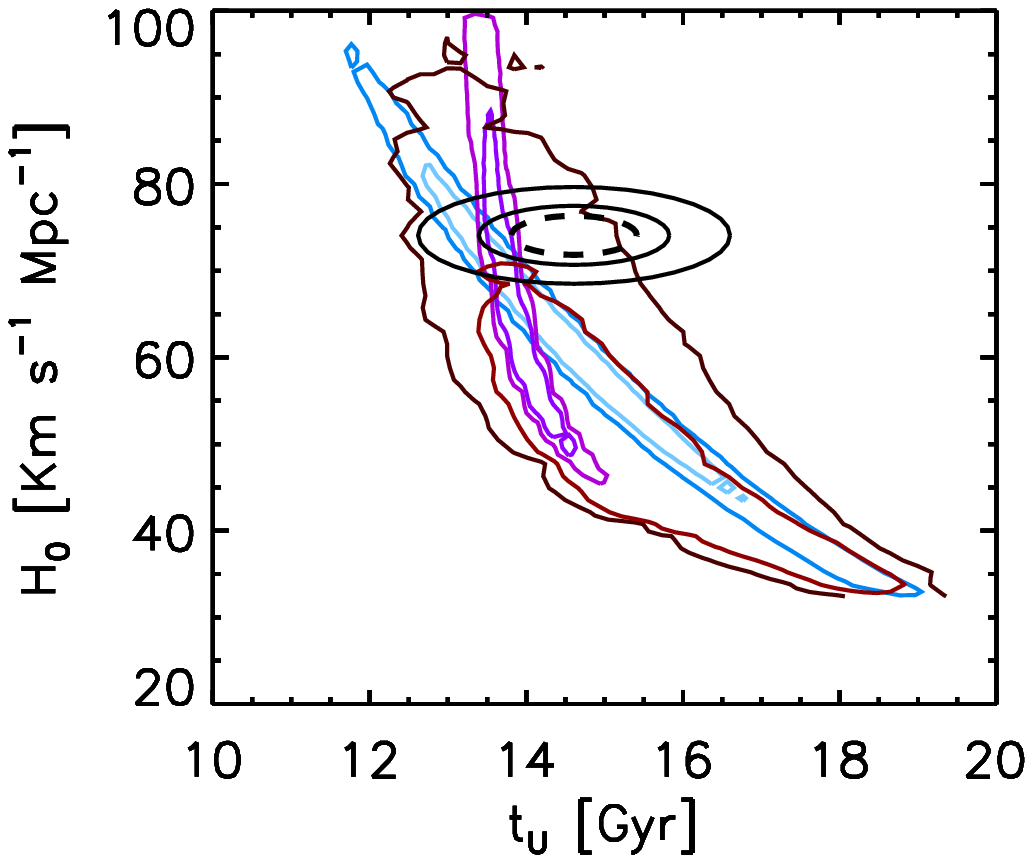}
                  \includegraphics[scale=0.55]{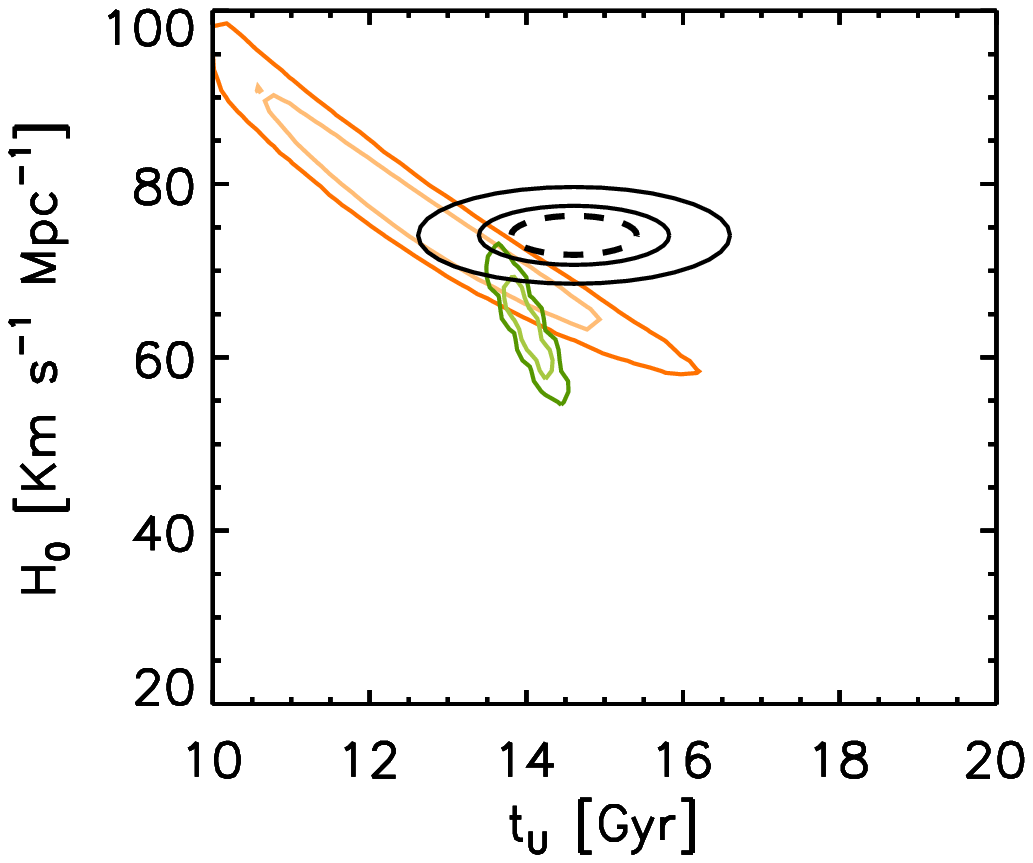}
        \caption{Comparison of WMAP9 constraints  in the $t_U$--$H_0$ plane for several models and direct $z=0$ measurements. The WMAP9 constraints are for  the following models. Left panel: Non-flat $\Lambda$CDM (blue), non flat wCDM (brown), flat wCDM (violet); Right panel: $N_{\rm eff} \Lambda$CDM (red-orange), $\Lambda$CDM with massive neutrinos  (green).
  For the local measurements   the dashed black line shows the 1-$\sigma$ (one parameter) and the solid black  lines show the 1 and 2 $\sigma$ joint  confidence regions.   In the two panels the axis ranges are the same for ease of comparison.      \label{fig:all}}
\end{figure}

Fig.~(\ref{fig:all}) shows the WMAP9 constraints in the $H_0$, $t_U$ plane for several simple extensions of the baseline  $\Lambda$CDM model (for which this particular parameter combination is most affected).  These models are: a $\Lambda$CDM model where the flatness assumption has been dropped,  a flat cold dark matter model where the dark energy equation of state parameter $w$ is not $-1$ but still constant (wCDM), a non flat wCDM model, a flat $\Lambda$CDM model where the effective number of neutrino species in not  the standard $3.04$ ($N_{\rm eff}\Lambda$CDM) and a  flat $\Lambda$CDM  where neutrinos are not massive.  We also show for comparison the local measurements. While local measurements are not competitive with CMB constraints for the baseline  flat $\Lambda$CDM  model, for many simple variations from this model the local measurements add  useful extra information.

\begin{figure}[h]

\hspace*{-0.7cm}
                \includegraphics[scale=0.42]{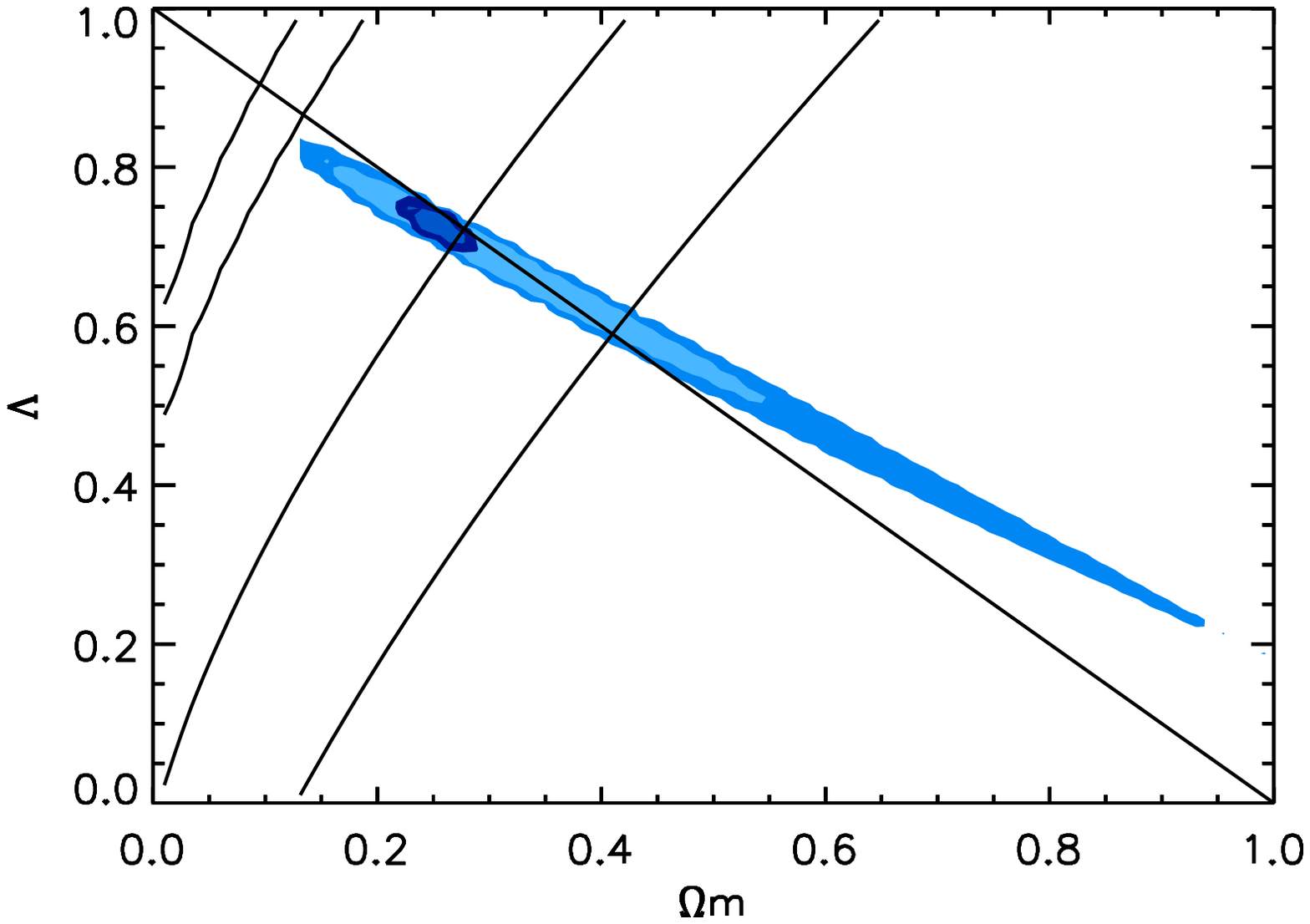}
                 \includegraphics[scale=0.42]{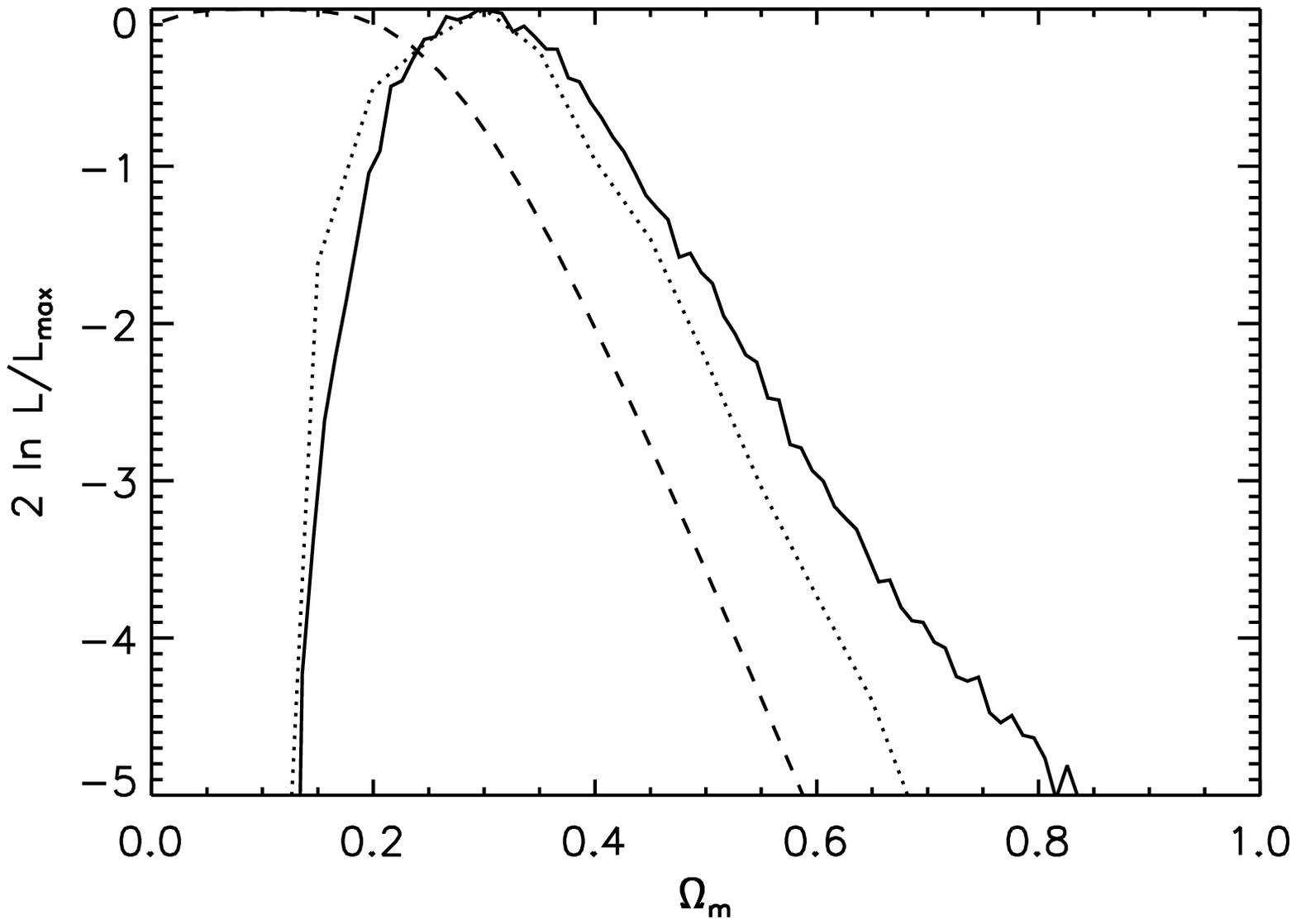}
        \caption{Left: Constraints (1- and 2 $\sigma$-joint) in the $\Omega_m$--$\Omega_\Lambda$ plane. The solid diagonal line shows the locus of spatially  flat models. Light blue: WMAP9. Transparent: Combination of $H_0$ and age measurements.  If the underlying model was truly incorrect there should be no reason for the  two sets of constraints to agree. Dark blue: total combination. When combining with CMB, the $H_0$ constraint is the one with most statistical power.  The dark blue contours are virtually indistinguishable from  those obtained from $WMAP9+H_0$ only.    Right: $\Omega_m$ constraints. For a non-flat $\Lambda$CDM model WMAP9 observations nicely bounds $\Omega_m$ from below (solid and dotted lines, same style as previous figure). $H_0$+$t_U$ measurements (dashed line) bound it from above.       \label{fig:omegamomegal}}
\end{figure}

We now consider what cosmological constraints can be obtained from local measurements.
We begin by relaxing the flatness assumption. 
Note that the combination of local $H_0$ and $t_U$ measurements already places constraints in the $\Omega_m$--$\Lambda$ plane. These constraints (shown as the transparent set of contours in the left panel of Fig.~(\ref{fig:omegamomegal})),  are not particularly tight but are nicely orthogonal to the  CMB ones (light blue shaded regions).  Note that the two local measurements alone favor a low density Universe. On the other hand,  if the underlying model was truly incorrect there should be no reason for the  two set of constraints (the high z and local ones) to agree.
It is interesting also to consider how WMAP9 and $H_0$+$t_U$ constrain the Universe  matter density in this extension of the baseline model: this is shown in the right panel of Fig.~(\ref{fig:omegamomegal}).

The addition of the curvature parameter lets models  with large $\Omega_m$ to still be a good fit to CMB data but bounds $\Omega_m$ from below. The combination of $H_0$ and $t_U$ on the other hand does not bound  $\Omega_m$ from below (there can be empty non-flat models that have  $H_0$ and $t_U$ in agreement with observations, but it bounds  $\Omega_m$ from above (too high matter density  would make the Universe too young  for an Hubble constant in agreement with observations), see Ref.~\cite{Hcosmo}.

\begin{figure}[h]
        \centering
                \includegraphics[scale=0.6]{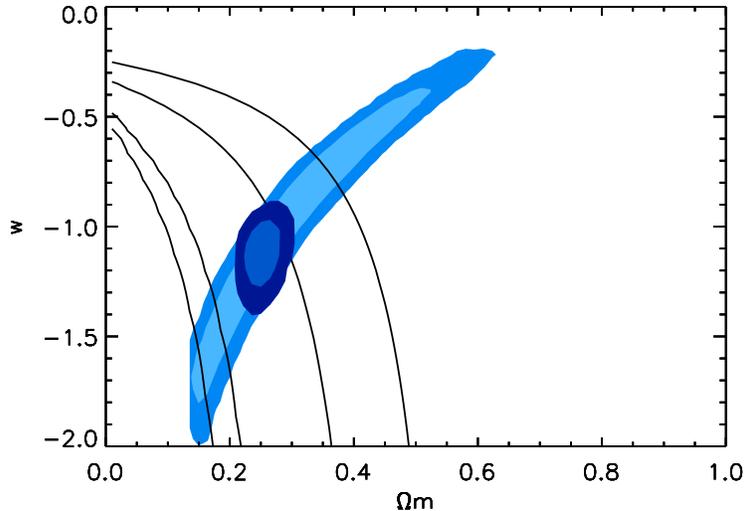}       
        \caption{Constraints (1- and 2 $\sigma$-joint) in the $\Omega_m$--$w$ plane. Light blue: WMAP9. Transparent: Combination of $H_0$ and age measurements.  If the underlying model was truly incorrect there should be no reason for the  two sets of constraints to agree. Dark blue: total combination. When combining with CMB, the $H_0$ constraint is the one with most statistical power. \label{fig:w}}
\end{figure}

We next consider a spatially flat Universe but where the equation of state parameter for dark energy is different from $w=-1$, but still constant.
The resulting constraints from WMAP9 and the combination $H_0$+$t_U$ are shown in Fig.~(\ref{fig:w}). Once again the local measurements  give  constraints that  are nicely orthogonal to CMB ones and the combination of these local measurements with CMB ones nicely break the degeneracy..

 In is well known that in models where the neutrino properties are changed from the standard  three massless species, CMB data alone yield degeneracies between neutrino properties and $H_0$ and $t_U$.  
 For example there has been a long debate in the literature of  wether cosmological data  favor (or not) extra effective species, which goes under the name of "dark radiation" (for a review see \cite{whitepaper} and references therein, for a bayesian model selection approach see \cite{feeney}).

\begin{figure}[h]
        \centering
                \includegraphics[scale=0.75]{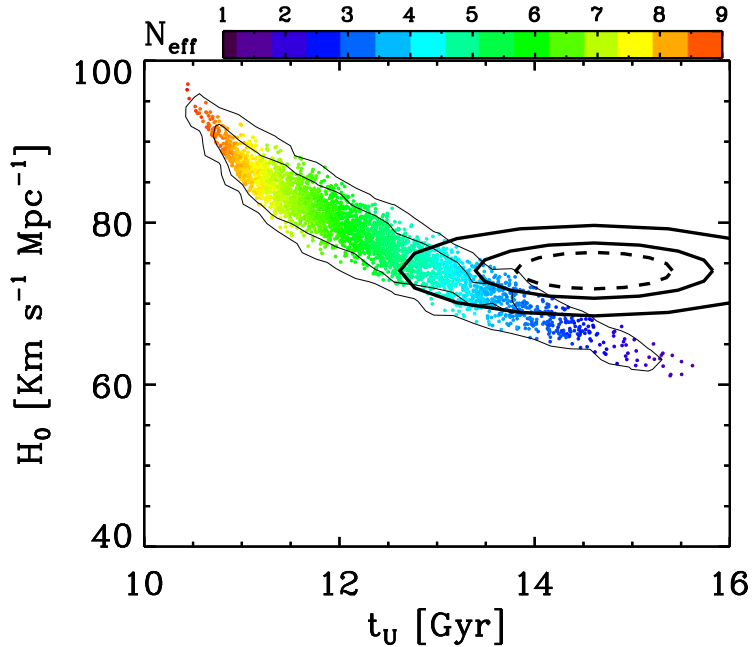}
               \caption{Constraints in the $t_U$--$H_0$ plane for a $N_{\rm eff}\Lambda$CDM model. The points are extracted from a WMAP7 Monte Carlo Markov Chain and are selected so that $2\ln(L/L_{max})<6.17$. They are color-coded by the corresponding value of $N_{\rm eff}$. The  thick ellipse shows the 1-$\sigma$(one parameter-dashed),  and 1 and 2 $\sigma$ (joint, solid)  constraints from the local measurements of $H_0$ and $t_U$. The thin lines  are the posterior  CMB 1 and 2 $\sigma$ joint confidence regions. The  ``edge" on the left hand side of the CMB degeneracy is due to the  hard prior adopted on $N_{\rm eff}$. \label{fig:ageHneff}
}
\end{figure}

Fig.~(\ref{fig:ageHneff}) illustrates this. It is a scatter plot in the $t_U$--$H_0$ plane; the points are extracted from a WMAP7 Monte Carlo Markov Chain and are selected so that $2\ln(L/L_{max})<6.17$. They are color-coded by the corresponding value of $N_{\rm eff}$. The  thick ellipse show the 1-$\sigma$(one parameter-dashed),  and 1 and 2 $\sigma$ (joint, solid)  constraints from the local measurements of $H_0$ and $t_U$. The thin lines  are the posterior  CMB 1 and 2 $\sigma$ joint confidence regions.
We have used a WMAP7 Monte Carlo Markov Chain rather than WMAP9 because for the WMAP7  chains we use the primordial helium abundance $Y_P$ was varied along with $N_{\rm eff}$ rather than being kept constant at nucleosynthesis value.
The faster cosmological expansion due to the neutrino background changes the acoustic and damping angular scales of the CMB, but  equivalent changes can be produced by varying  the primordial helium abundance e.g., \cite{Bashinsky}. A direct comparison of Figs.   (\ref{fig:ageHneff}) and (\ref{fig:all})  clearly shows that the choice of WMAP7 or WMAP9 does not really affect  the gist of our argument.

 We find that WMAP+$t_U$ yields $N_{\rm eff}=2.58 \pm 0.75$ while WMAP+$H_0$+$t_U$ yields  $N_{\rm eff} = 3.53\pm 0.54$ (recall that WMAP+$H_0$ yields $N_{\rm eff}=4.31 \pm 0.73$).
It is apparent from Fig.~(\ref{fig:ageHneff}) that the $H_0$ measurement tends to cut the CMB degeneracy   centered around $N_{\rm eff}=4 \sim 5$ while the $t_U$ measurement is centered around  $N_{\rm eff}=3$. This is a consequence of the ``hint" of tension mentioned earlier. Clearly, tightening the errors on the age measurement would greatly help.    We will return to this point in \S \ref{sec:tension}. 
 The addition of $N_{\rm eff}$ as a parameter does not seem to bring in better agreement CMB and local measurements (although, as discuss before the hint of tension is not statistically significant).
 
 Here we note that the fact that  CMB data for $N_{\rm eff}$ significantly higher than 3,  predict an age of the Universe which is low compared with the estimated ages of the oldest objects (e.g, globular clusters)  has been  mentioned before in the literature e.g., \cite{deBernardis:2007bu, Kristiansen/Elgaroy:11, Archidiacono:2011gq}.
 
\begin{figure}[h]
        \centering
                \includegraphics[scale=0.75]{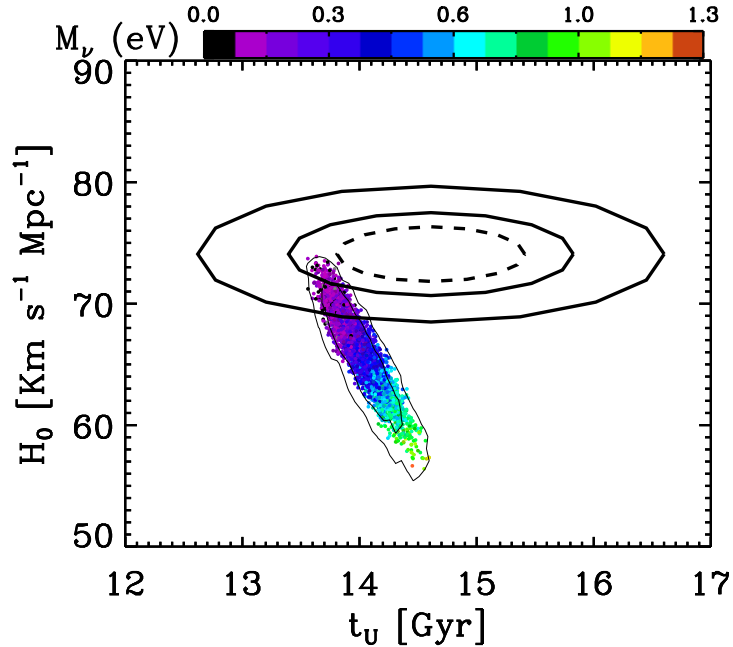}
               \caption{Constraints in the $t_U$--$H_0$ plane for a $N_{\rm eff}\Lambda$CDM model. The points are extracted from a WMAP7 Monte Carlo Markov Chain and are selected so that $2\ln(L/L_{max})<6.17$. They are color-coded by the corresponding value of $M_{\nu}$. The  thick ellipse shows the $1-\sigma$ (one parameter-dashed),  and 1 and $2-\sigma$ (joint, solid)  constraints from the local measurements of $H_0$ and $t_U$. The thin lines  are the posterior  CMB 1 and $2-\sigma$ joint confidence regions.  \label{fig:ageHmnu}
}
\end{figure}

While the $t_U$ determination greatly helps constraining the number of effective species $N_{\rm eff}$, the $H_0$ measurement help tightening the  neutrino masses constraint (which is not too affected by the age determination). This can be seen from the right panel of  Fig.~(\ref{fig:all}), where the green contours correspond to the $\Lambda$CDM model with the addition of massive neutrinos. The error on the age measurement and the orientation and size of the CMB degeneracy  mean that the  $t_U$ determination does not help with tightening the neutrino mass constraint. 
This is further illustrated in Fig.~(\ref{fig:ageHmnu}).

As for Fig.~(\ref{fig:ageHneff}) it is a scatter plot in the $t_U$--$H_0$ plane; the points are extracted from a WMAP7 Monte Carlo Markov Chain and are selected so that $2\ln(L/L_{max})<6.17$. They are color-coded by the corresponding value of $M_{\nu}$.  The  thick ellipse show the $1-\sigma$(one parameter-dashed),  and 1 and $2-\sigma$ (joint, solid)  constraints from the local measurements of $H_0$ and $t_U$. The thin lines  are the posterior  CMB 1 and $2-\sigma$ joint confidence regions. 
It is clear that if the central values of $H_0$ for CMB and the direct measurement  coincided,  the limit on $M_{\nu}$ from the CMB+$H_0$ combination would not have been  so tight.  

 \begin{figure}[h]
 \hspace*{-0.7cm}
                \includegraphics[scale=0.65]{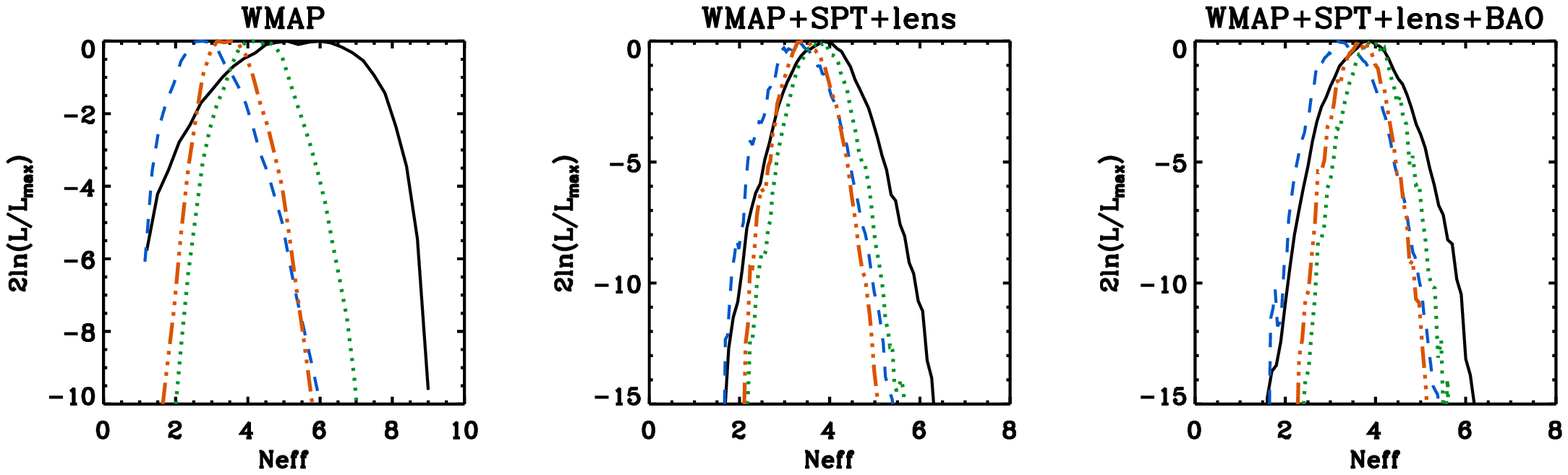}
                \hspace*{-0.7cm}
                \includegraphics[scale=0.62]{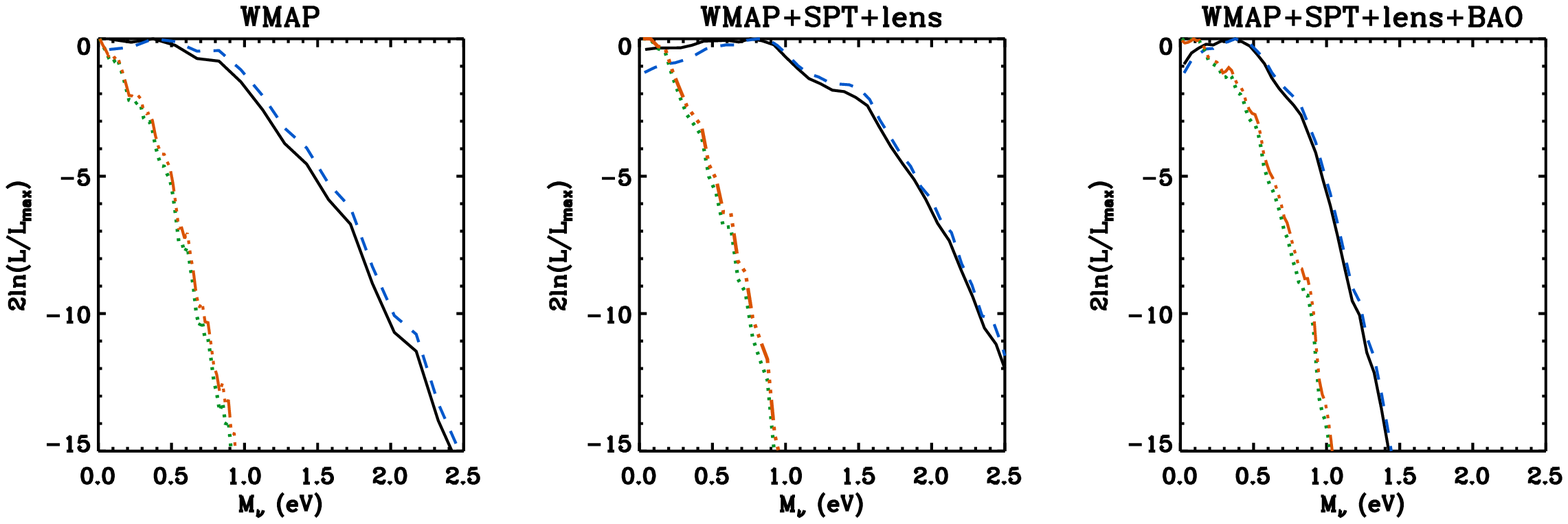}
               \caption{Constraints on neutrino properties obtained by adding one extra parameter to the baseline $\Lambda$CDM model, the parameters are the effective number of species $N_{\rm eff}$ (top) and the total mass $M_{\nu}$ (bottom). The black (solid) lines  correspond to the data combination reported in the panel title  without age and $H_0$. Dashed (blue) lines show the effect of including the $t_U$ constraints; dotted (green) lines   show the effect of including the $H_0$ constraints and dot-dot-dot-dashed (orange) lines  show the effect of including the $H_0$  and $t_U$ constraints. \label{fig:neffmnuageH1d}}
\end{figure}

The resulting marginalized constraint on the neutrino masses are shown in Fig.~\ref{fig:neffmnuageH1d}). In this case for completeness we have considered not only WMAP7  but also WMAP7 data  in combination with  the South Pole Telescope (SPT) determination of the damping tail  of \cite{SPT} and their constraints on the lensing defection power spectrum from \cite{SPTLens} as indicated in the panel's titles.  The black (solid) lines  correspond to the data combination without age and $H_0$. Dashed (blue) lines show the effect of including the $t_U$ constraints; dotted (green) lines   show the effect of including the $H_0$ constraints and dot-dot-dot-dashed (orange) lines  show the effect of including the $H_0$  and $t_U$ constraints.

\section{Tension?}
\label{sec:tension}
We return here on the hint of tension discussed above. Recently, tension between different data sets (or actually between  CMB constraints on parameters or parameter combinations obtained   within  a given model and the value measured by other data-sets) has been found. This is discussed extensively in Ref.~\cite{Hou/etal:12}. There they use WMAP7 and SPT data, and compare the CMB constraints on derived parameters such as $H_0$ and  the Baryon Acoustic Oscillation (BAO) measurement on the angle-averaged location of the acoustic peak. This is  effectively a  constraint on a weighted average of the  Universe expansion history between $z=0$ and $z=0.57$.  Similarly to what we find here,  they find that, taken individually,  the  expansion history measurements show no tension with the CMB but taken together they do. 
 
 Relevant to the discussion presented here, they find that CMB+$H_0$ favors larger $N_{\rm eff}$ than CMB+BAO; the latter is in line with our $t_U$ constraints.
 
 They also find that for all the models they considered,  no single parameter extension to the baseline $\Lambda$CDM model helps to  alleviate this tension.

Given the  relatively large errors on $t_U$ and the fact that Ref.~\cite{Hou/etal:12} focussed on the WMAP7+SPT combination while here we use WMAP9 only,  the ``hint" of tension we find here is much less statistically significant than what  Ref.~\cite{Hou/etal:12}  finds.
For the purpose of the present work, it is important to bear in mind that 1-$\sigma$ error means that one  out of three ``events" should fall outside the reported margin. Since here we consider three different data-sets, it is reasonable to expect that  for one of the three  the true value falls outside the 1-$\sigma$ error. 
 However it is interesting to entertain the possibility that the tension is real and  some  systematic effect of new physics is behind. We have  so far discussed the second possibility. It is worth to keep in mind that the $H_0$ measurement could be affected by cosmic variance, see Ref.~\cite{marra},   which could ease the tension. If this systematic effect is to blame, possibly an improved  analysis of the local velocity field e.g., \cite{Neill}, could be used  to correct for it.

\section{Future prospects}
\label{sec:future}

A precision  of the  $H_0$  measurement that approaches 1\%  could be feasible \cite{Riess/etal:2011} and it is the goal of the SH$_0$ES  project to reach the 2\% error on $H_0$.
As \% precision seems to be the goal for precision cosmology 
we can think of what would be required to obtain an age  determination of $\approx 1-2\%$.
Table~1 in \cite{Bond} details the error breakdown and shows that  largest contributors to the error budget of the age of the stars are oxygen abundance, effective temperature determination, distance and photometry, in decreasing order of importance.
 The  HD 140283-derived distance and photometry errors  can be reduced by observing many similar stars and better determining their distances, as will be the case of the GAIA\footnote{{\tt http://sci.esa.int/science-e/www/area/index.cfm?fareaid=26}} mission (launch date October 2013), which will be able to determine distances to such bright stars with at least an order of magnitude better accuracy and many more of them.  The main source of error at present  however is due to the determination of surface abundances (especially $[O/H]$) and effective temperature, which  can be decreased by doing better spectroscopic observations with current 10m class telescopes and better 3D theoretical modelling (e.g., \cite{NordlundI, NordlundII}).  Note that the dominant source of --systematic--error, $[O/H]$, arises as the error is estimated from the dispersion among measurements of different authors and of different indicators (set of lines in the stellar  spectra). Each of these measurements has a  much smaller statistical error.   It is believed that the forbidden $[O_{I}]$ line at $6300 \AA$ line is the most reliable indicator for oxygen abundances, as oxygen abundance so derived  is not sensitive to stellar parameters. Its uncertainty is dominated by the error in equivalent width; but the difficulty is that this line  is very weak in  sub-giant stars and could be blended with other lines. One possible avenue therefore would be to observe this indicator with higher signal-to-noise and  therefore avoid relying on determinations from other --more model-dependent-- indicators. 

If the above errors could be reduced,  then the next most important contribution would arise from the uncertainty in the solar oxygen abundance, that is, to finally settle on the actual value of oxygen in the solar photosphere by 3D modelling of the Sun \cite{Sun}. 

Finally, note that systematic uncertainties coming from the theoretical modelling of stellar evolution itself (like equation of state, nuclear reaction rates or opacities) are very small, accounting for changes in the age of $\sim 1 $\%. 

If the age errors could be reduced from 5\% to 2\% (comparable with current statistical errors only) the uncertainty over formation time would still be largely subdominant. However such reduced error would make the age measurement really competitive.
In  Figs.~(\ref{fig:2dcomparison},\ref{fig:all},\ref{fig:ageHneff},\ref{fig:ageHmnu}), the 2-sigma joint confidence  region will  approximately coincide with the dashed line.

Therefore, if the central value remained the same, the hint of tension between CMB and local measurements $H_0$, $t_U$ would become a clear detection.

If instead  the central value moved and there were no tension,  a much reduced error on cosmological parameters would be obtained. For example the (1$\sigma$ error  on $N_{\rm eff}$ from WMAP+$t_U$ would  be $\pm 0.3$ which would clearly (i.e. at about $3-\sigma$level) rule out (or in) sterile neutrinos; the  error from the WMAP+$t_U$+$H_0$ combination  would not change significantly.

\section{Discussion and Conclusions}

Local, cosmology-independent  measurements  of cosmologically relevant quantities can be used to test self consistency of the currently favored cosmological model  and to constrain deviations from it.  There are not many cosmology-independent determinations of cosmologically interesting parameters. Here we have focussed on {\it local} measures, which happen to be only testing the Universe expansion history: $H_0$ and age. The advantage of using {\it local} measurements is that in all cosmological models parameters are defined at $z=0$ and thus no cosmology-dependent extrapolation is needed. It is worth however noting that  direct measurements of e.g. $H(z)$, through  (relative) radial  Baryon acoustic oscillations (BAO)  or spectroscopic dating of old galaxies also offer  cosmology independent constraints. These however are at $z>0$ and thus not a focus of this paper.
While we have examined only a small sample of deviations from the baseline $\Lambda$CDM model it is worth  to note that the combination of local $H_0$ and age measurements can be crucial to constrain models where the dark energy equation of state is not constant, where there is a coupling between dark matter and dark energy or where dark energy or dark matter decay, etc.
The fact that  high-precision, local, cosmology-independent  measurements are so far only available for measures of the expansion history makes it difficult to test in a model-independent way for deviations for general relativity (or rather such deviations could be described  by a non constant dark energy model).

In addition to the Hubble constant and the age of the universe, a third quantity of interest to measure locally  would be the current matter density. This measurement has not been explored systematically as the two others and there is not a real measurement with accurate and precise errors, both systematic and statistical (the available measurement being more than a decade old \cite{bahcall}). In order to do this independent of the cosmological model, one would have to use, for example,  a measurement of the mass-to-light ratio in nearby clusters, some difficulties have been highlighted recently by Ref.~\cite{tinker}, but they explore a road forward to obtain reliable measurements without a dependence on a cosmological model.

To conclude,  the main aim of this paper was to take into account the new local, direct measurements for  the age of the universe along with recent, local,  Hubble constant measurements, which have reached an accuracy of better, or at the level of 5\%, even including systematic uncertainties. We have pointed out that these measurements, as they are cosmology-independent, provide the means to test cosmology without relying on a particular model that thus implies testing its parameters. 

With this in mind, we  have explored quantitatively with recent CMB data, what constraints one could put on the current $\Lambda$CDM model, i.e. are the high-redshift  and low-redshift Universes  consistent (within a given model)? We found that with the current precision of local measurements they are. However, there is a slight hint of tension that could be further investigated by decreasing the error in the local measurements by a factor of two. This could elucidate if there is some physics beyond the current $\Lambda$CDM model.

Given that the current ``hint" of tension does not seem significant, the current local data can be used to put constraints on extensions of the $\Lambda$CDM model. We have shown how they provide nearly orthogonal constraints on the geometry of the universe and the equation of state of dark energy to the CMB ones. Of particular interest is the constraints we obtain when the $\Lambda$CDM model is extended to include extra number of neutrinos. We have shown how current constraint help enormously at constraining this particular extension of the model, in particular when local measurements on the age are added $N_{\rm eft} < 4$ at the 95\% level, in good agreement with an independent measurement of $N_{\rm eff} = 3 \pm 0.5$ by \cite{cooke} using a measurement of primordial deuterium.

While there are on-going programs to systematically reduce and control systematic and statistical errors in the determination of $H_0$, as far as we know  there is no  such program for a local direct  determination of the age of the Universe.
It is interesting that the GAIA mission will reduce the uncertainty in the ages of old stars due to distance to below the level due to the chemical abundance uncertainty. On the other hand, for the globular clusters GAIA should deliver parallaxes with an accuracy of few tens of micro arcseconds, i.e. orders of magnitude better than current limits.  So for both  techniques (using galactic stars or globular clusters) in the error budget of  the age  determination the dominant quantity will be the chemical abundance of the stars.
This will need to be improved by better and targeted observations by 10-30m class telescopes, but it is interesting that a targeted program on this front could bring robust ages to the \% level.

We argue that reducing the error budget of local measurement  of the age of the Universe  at the $\approx \%$ level would provide us the means to start exploring cosmology from a model independent perspective and move beyond the very successful parameter estimation within a model.\\

{\bf Acknowledgments:} LV   is supported by European Research Council under the European CommunityÕs Seventh Framework Programme grant FP7-IDEAS-Phys.LSS . LV and RJ acknowledge Mineco grant FPA2011-29678- C02-02. SMF is supported by STFC.  SMF is supported by STFC and a grant from the Foundational Questions Institute (FQXi) Fund, a donor-advised fund of the Silicon Valley Community Foundation on the basis of proposal FQXiRFP3-1015 to the Foundational Questions Institute. We acknowledge the use of the Legacy Archive for Microwave Background Data Analysis (LAMBDA). Support for LAMBDA is provided by the NASA Office of Space Science.
\bibliographystyle{model1-num-names}


\end{document}